\documentclass[preprint,aps]{revtex4}
\usepackage{epsfig,amsmath,amssymb}
\newcommand{\be}{\begin{equation}}
\newcommand{\ee}{\end{equation}}

\newcommand{\bit}{\begin{itemize}}
\newcommand{\eit}{\end{itemize}}
\newcommand{\bea}{\begin{eqnarray}}
\newcommand{\eea}{\end{eqnarray}}

\newcommand{\kagome}{{ kagom\'e }}

\newcommand{\Kagome}{{\Kagome} }

\begin{document}

\title{
Localized-magnon states in strongly frustrated quantum spin lattices
}

\author{J. Richter}
\affiliation{Institut f\"ur Theoretische Physik, Universit\"at Magdeburg,
      P.O. Box 4120, D-39016 Magdeburg, Germany}
\date{\today}

\begin{abstract}
Recent developments concerning  localized-magnon eigenstates in 
strongly frustrated spin lattices and their effect on the  
low-temperature physics of
these systems in high magnetic fields are reviewed.
After illustrating the construction and the properties 
of localized-magnon states we describe the plateau and the
jump in the magnetization process caused by these states.
Considering appropriate lattice deformations fitting to the localized magnons 
we discuss a spin-Peierls instability in high magnetic fields related to
these states.
Last but not least we consider the degeneracy of the localized-magnon
eigenstates and the related  thermodynamics in high
magnetic fields. In particular, 
we discuss  the
low-temperature maximum in the isothermal entropy versus field curve and the    
resulting enhanced magnetocaloric effect, which allows efficient 
magnetic cooling from quite large temperatures down to very low ones. 
\end{abstract}


\maketitle

\section{Introduction}
The interest in quantum spin
antiferromagnetism has a very long tradition, see e.g. Ref. 
\onlinecite{mattis81}.
Stimulated by the recent progress in synthesizing
 magnetic materials with strong quantum fluctuations
\cite{lemmens}
particular attention has been paid on low-dimensional  
quantum magnets showing novel quantum phenomena like spin liquid
phases, quantum phase transitions or  plateaus and jumps in the magnetization
process.
However, quantum spin systems are
 of interest in their own right  as examples of strongly interacting
 quantum  many-body systems.

We know from the Mermin-Wagner theorem \cite{mermin66dd} that 
 thermal fluctuations are strong enough to
 destroy magnetic long-range order (LRO) for Heisenberg spin systems in
dimension $D<3$ at any finite temperature $T$. 
For $T=0$, where  only quantum fluctuations are present, the situation seems
to be more complicated. While for one-dimensional (1D) antiferromagnets, in
general, the quantum fluctuations are strong enough to prevent magnetic
LRO,  the competition between interactions and fluctuations is
well balanced in two dimensions and one meets magnetic LRO as
well as magnetic disorder at $T=0$ in dependence on details of the lattice
\cite{lhuillier01,moessner01,lhuillier03,Richter04,hon04}.  
It was pointed out  many years ago by Anderson and Fazekas
 \cite{anderson73dd,fazekas74} that competition of
magnetic bonds for instances due to triangular configurations of
antiferromagnetically interacting spins
may influence this balance and can lead to disordered ground-state
phases in two-dimensional (2D) quantum antiferromagnets. 
In the context of spin glasses
this competition of bonds, 
later on called frustration, was discussed in great detail.
These studies on spin glasses 
have shown that frustration   
may have an enormous influence 
on ground state and thermodynamic properties \cite{binder86}  of spin systems.

The investigation of frustration effects in  spin systems,
 especially in combination 
with strong quantum fluctuations,  is currently
a 'hot topic' in solid state physics.
We mention some
interesting features like quantum disorder, incommensurate spiral
phases, {\it 'order by disorder'} phenomena to name a few, which might appear in 
frustrated systems.
The theoretical study of frustrated quantum 
spin systems is challenging and is often faced with particular
problems. While for unfrustrated systems a wide class of well 
developed many-body
methods are available, at least some of them, e.g. the powerful Quantum
Monte Carlo Method, fail  
for frustrated  systems.
Furthermore several important exact statements like the Marshall-Peierls
sign rule \cite{marsh55} and the Lieb-Mattis theorem \cite{lieb62}
are not generally valid  if frustration is present (see e.g.
\cite{ri_epl94,ri_jmmm95}).

On the other hand, the investigation of strongly frustrated magnetic 
systems surprisingly led
to the discovery of several new exact eigenstates.
To find 
exact eigenstates of quantum many-body systems is in general a rare
exception. For spin systems one has 
only a few examples. The simplest example for an exact eigenstate
is the 
fully polarized ferromagnetic state, which becomes the ground state of an
antiferromagnet in a strong magnetic field.
Furthermore the one- and two-magnon excitations above the 
fully polarized ferromagnetic state  also can be
calculated exactly (see, e.g. \cite{mattis81}). 
An  example for non-trivial eigenstates is Bethe's famous 
solution for the 1D 
Heisenberg antiferromagnet (HAFM)
\cite{bethe}. 
Some of the eigenstates 
found for  frustrated quantum magnets  
are  of quite simple nature and their physical properties, 
e.g. the spin correlation functions, can be calculated analytically.
Note that such states are often eigenstates of the unfrustrated system, too,
but they are irrelevant for the physics of the unfrustrated system 
if they are lying somewhere in the spectrum.
However, the interest in these
eigenstates   comes from the fact that  they may 
become ground states
for particular values of frustration.
Therefore these exact eigenstates play an important role either as 
ground states of real quantum magnets or at least as ground states
of idealized models which can be used as reference states
for more complex quantum spin systems.

Two well-known examples for simple eigenstates 
of strongly frustrated
quantum spin systems are the Majumdar-Gosh state of the 1D 
$J_1-J_2$ spin-half HAFM \cite{majumdar}  
and the orthogonal dimer state of the
Shastry-Sutherland  model \cite{shastry81}.  Both eigenstates are 
product states built by dimer singlets. They 
become ground states only for strong 
frustration. These eigenstates indeed play
a role in realistic materials. The Majumdar-Ghosh state has some
relevance in quasi-1D spin-Peierls materials like $CuGeO_3$ (see e.g. 
\cite{cugeo}).
The orthogonal dimer state of the Shastry-Sutherland model 
is the magnetic ground state of the quasi-two-dimensional 
SrCu$_2$(BO$_3$)$_2$ compound \cite{srcubo}. 
Other frustrated spin models in one, two or three 
dimensions are known which have also dimer-singlet product states as
ground states (see e.g. \cite{pimpinelli,ivanov97,japaner3d, schmidt}).
Note that   
these dimer-singlet product ground states  have gapped magnetic excitations
and lead therefore to  
a plateau in the magnetization $m$ at $m=0$.
Recently it has been demonstrated for the 1D counterpart of the
Shastry-Sutherland model
\cite{ivanov97,koga,schul02,schul02a}, that more general product eigenstates
containing chain fragments of finite length can 
lead to an infinite series of
magnetization plateaus \cite{schul02}.
Finally, we mention the so-called frustrated Heisenberg star where also exact
statements on the ground state are known \cite{starI}.

In this paper we review recent results 
concerning a new class of exact eigenstates appearing in
strongly frustrated antiferromagnets, namely the so called localized-magnon
states. These states have been detected as ground states  of
certain frustrated antiferromagnets \cite{schn01,prl02,ri04} in a magnetic 
field and their 
relevance for 
physical properties of a wide class of frustrated magnets has been 
discussed in  
\cite{Richter04,schn01,prl02,ri04,sp04,zhito04,zhito04a,der04,star04,sqago04}.

\section{Localized magnon states} \label{states}
\begin{figure}
\begin{center}\epsfxsize=26pc
\epsfbox{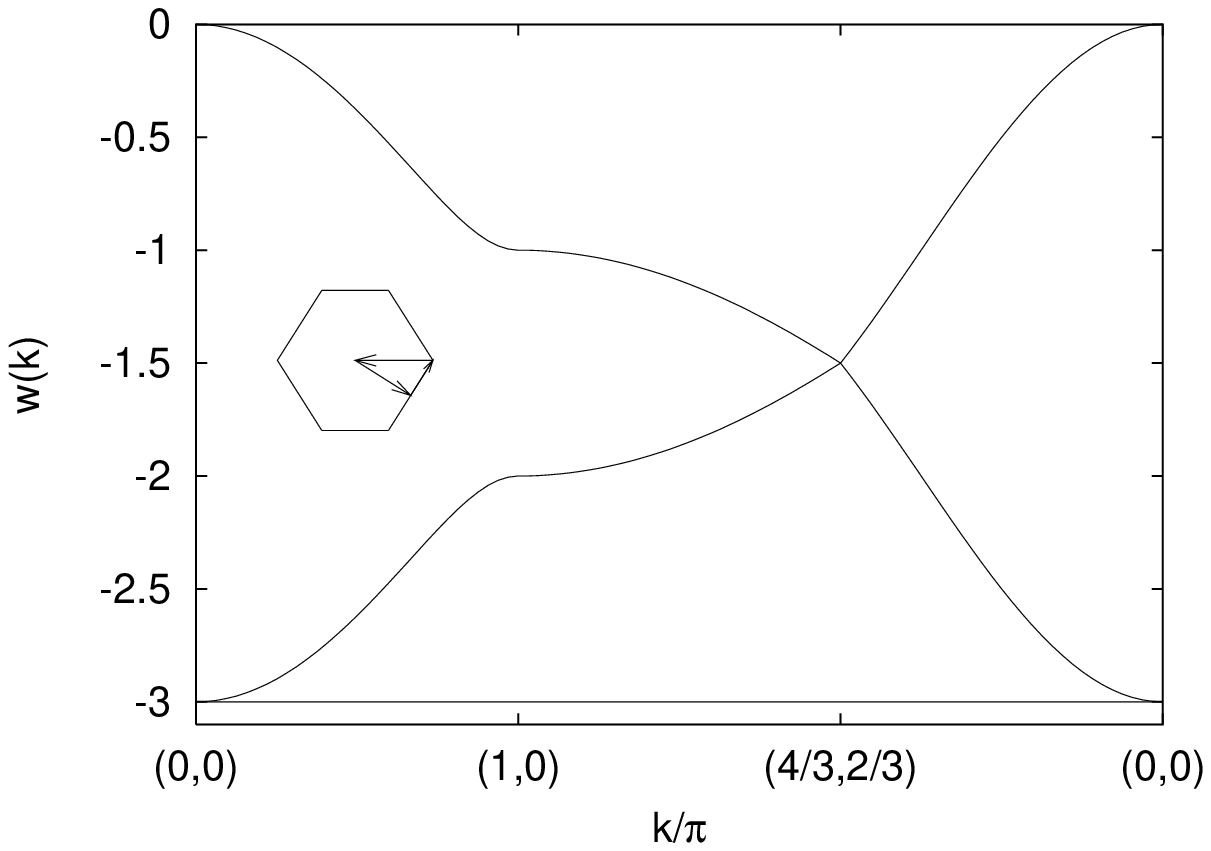}
\epsfxsize=4.5pc
\hspace{-2.9cm}\epsfbox{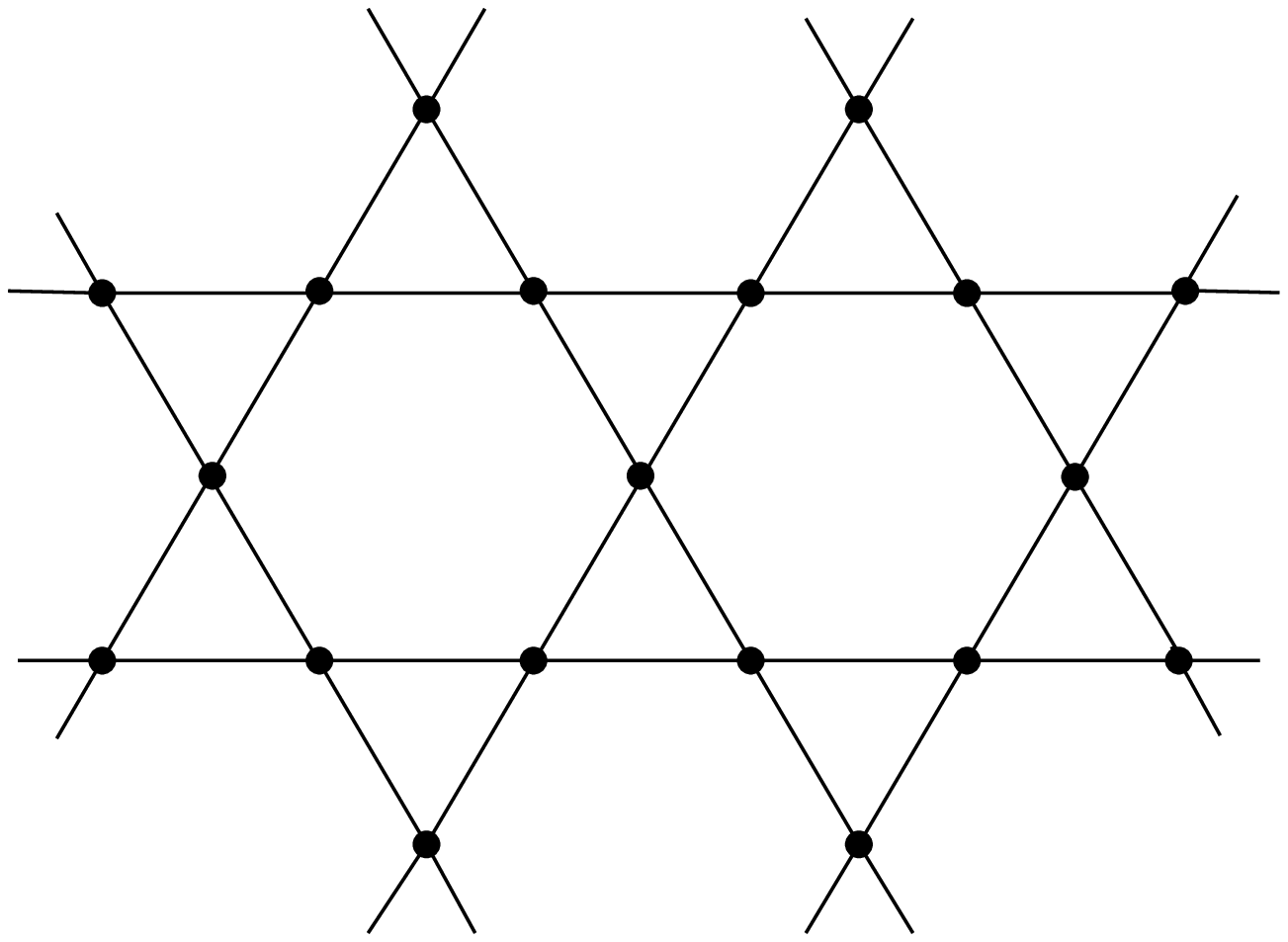}
\end{center}
\caption{\label{fig1} Excitation energies $w_i({\bf k})$
of the one-magnon states  for the isotropic spin-half 
Heisenberg antiferromagnet
(i.e.
$J=1$, $\Delta=1$, $s=1/2$ and $h=0$ in Eq. (\ref{eq1}))
on the \kagome  lattice (right inset). 
The left inset shows the path
in the Brillouin zone corresponding to k values used as x coordinate.
This figure was provided by J.~Schulenburg.}
\end{figure}

 We consider 
a Heisenberg $XXZ$ antiferromagnet for general spin quantum number $s$
in a magnetic field $h$
 \be \label{eq1}
 \hat{H}= \sum_{\langle ij \rangle} J_{ij}
\left\{\Delta \hat{S}_i^z \hat{S}_j^z + {1 \over 2} \left(
\hat{S}_i^{+} \hat{S}_j^{-} + \hat{S}_i^{-} \hat{S}_j^{+} \right) \right\}
       - h \hat{S}^z \; , \; J_{ij} \ge 0. 
\ee
The 
magnetization $M=\hat{S}^z=\sum_i \hat{S}_i^z$ 
commutes with
the Hamiltonian and is used as a relevant
quantum number to characterize the eigenstates of $\hat{H}$.
Let us consider strong magnetic fields exceeding the saturation field
$h_{sat}$. Then the system is in the fully polarized ferromagnetic  
eigenstate
 $|FM\rangle = 
\;  |+s , +s, +s , +s, +s \dots \rangle$, 
which will be  considered as the magnon vacuum state,
i.e. 
$|0\rangle =|FM\rangle$.
The one-magnon states above this vacuum are given by  
\be \label{eq2}
 |1\rangle =\frac{1}{c} \sum_i^N a_i\hat S_i^-|0\rangle 
\quad ; \quad \left( \hat{H} -E_{FM}\right)|1\rangle =  w_i({\bf k}) |1\rangle ,
\ee
where $E_{FM}$ ist the energy of the fully polarized ferromagnetic  
state
 $|FM\rangle $ and $c$ is a normalization constant. 
For several strongly frustrated lattices 
one observes flat dispersion modes   $w_i({\bf k})=const.$ of the lowest branch. 
One example is shown in
Fig.~\ref{fig1}.
Here we mention that there is some relation to the flat-band 
ferromagnetism in electronic
systems discussed by Mielke and Tasaki, see e.g. 
\cite{Mielke91b,92e,93d}.

\begin{figure}
\begin{center}\epsfxsize=22pc
\hspace{-2cm}\epsfbox{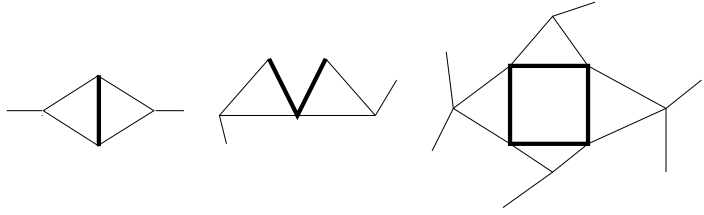}
\end{center}
\caption{\label{fig2} Typical lattice geometries supporting the 
localized-magnon states (\ref{eq3}). The magnon lives on the restricted area 
indicated by a thick line. 
}
\end{figure}

Consequently, one can construct localized sates by an appropriate 
superposition of extended states with different ${\bf k}$ vectors. 
The general form
of these localized states can be written as \cite{prl02,ri04}
\be \label{eq3}
|1\rangle_L
 =\frac{1}{c} \sum_i^N a_i\hat S_i^-|0\rangle
 =|\Psi_L\rangle|\Psi_R\rangle, 
 \hspace{0.4cm}  a_i \  \left\{ \begin{array}{cc}
    \neq 0 \;\: \forall i\in L & (local)\\
       = 0 \;\; \forall i\in R & (remainder)
   \end{array} \right.
\ee
where 
$|\Psi_L\rangle$  belongs to the localized excitation living on the local
region $L$  
and $ |\Psi_R\rangle$ describes  the fully polarized ferromagnetic 
remainder. We split the Hamiltonian  into three parts 
$\hat{H} = \hat{H}_{L} + \hat{H}_{L-R} + \hat{H}_{R}$, where $\hat{H}_{L}$
contains all bonds $J_{il}$
with $i,l \in L$,  $\hat{H}_{R}$
contains all bonds $J_{kj}$
with $k,j \in R$ and   $\hat{H}_{L-R}$
contains all bonds $J_{lk}$
with $l \in L$ and $k \in R$. 
The requirement that the localized-magnon state $|1\rangle_L$ is
simultaneously an eigenstate of all three parts of the Hamiltonian, i.e.   
$\hat{H}_{L} |1\rangle_L=e_L|1\rangle_L \;$,
$\hat{H}_{R} |1\rangle_L=e_R|1\rangle_L \;$
and $\hat{H}_{L-R} |1\rangle_L=e_{L-R}|1\rangle_L$,
leads to two criteria for the exchange bonds $J_{ij}$
\cite{ri04}, namely 
\be \label{eq4}
\sum_{l \in L} J_{rl} a_l = 0 \quad \forall \; r \in R  
\ee
and \be \label{eq5}
\sum_{r \in R} J_{rl} = const. \quad \forall \; l \in L.  
\ee
Eq. (\ref{eq4}) represents a condition on the bond geometry, whereas
Eq. (\ref{eq5}) is a condition for the bond strengths and 
is automatically 
fulfilled in uniform lattices with equivalent sites.
Note, however, that  the second condition is not a necessary one, i.e. one can
find models  with eigenstates of form (\ref{eq3}) violating (\ref{eq5}), see 
\cite{prl02}.  
This more general case appears if 
$| \Psi_L\rangle| \Psi_R\rangle$ is not an individual 
eigenstate of both $\hat H_L$
{\bf and} $\hat H_{L-R}$ but of $(\hat H_{L-R}+\hat H_{L})$. 
Hence, the geometry condition (\ref{eq4}) is the criterion of major
importance. A typical geometry fulfilling condition (\ref{eq4})
is realized 
by an even polygon surrounded by isosceles triangles, see Fig.~\ref{fig2}.
The lowest one-magnon state living on an even polygon has  
 coefficients $a_i$ in $|1\rangle_L$ 
alternating in sign. 
But also finite strings of two or three sites attached by appropriate triangles 
can fulfill the criterion (\ref{eq4}), see Fig.~\ref{fig2}.

\begin{figure}
\begin{center}
{\epsfig{file=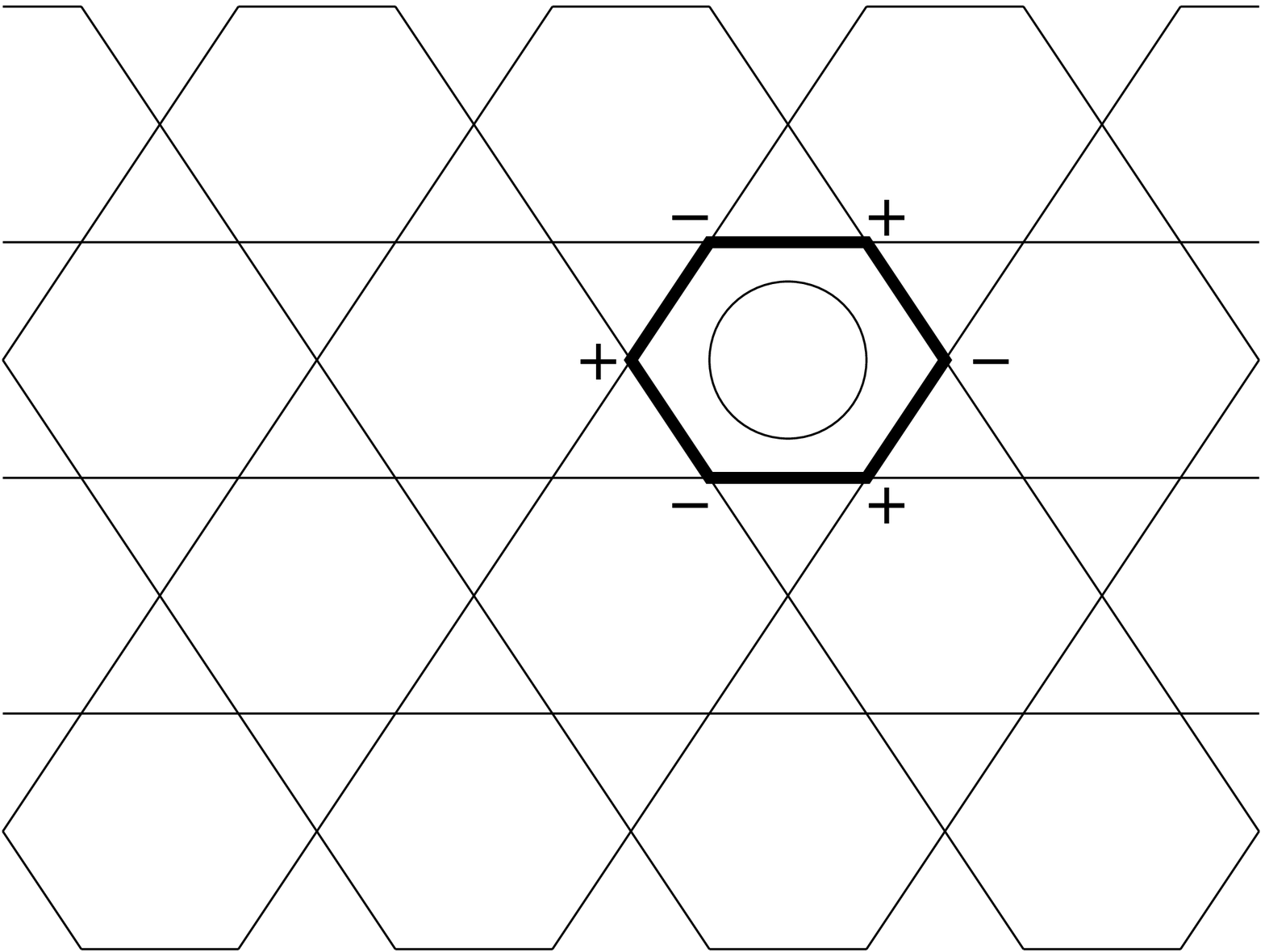,scale=0.21}} \hspace{0.5cm}
{\epsfig{file=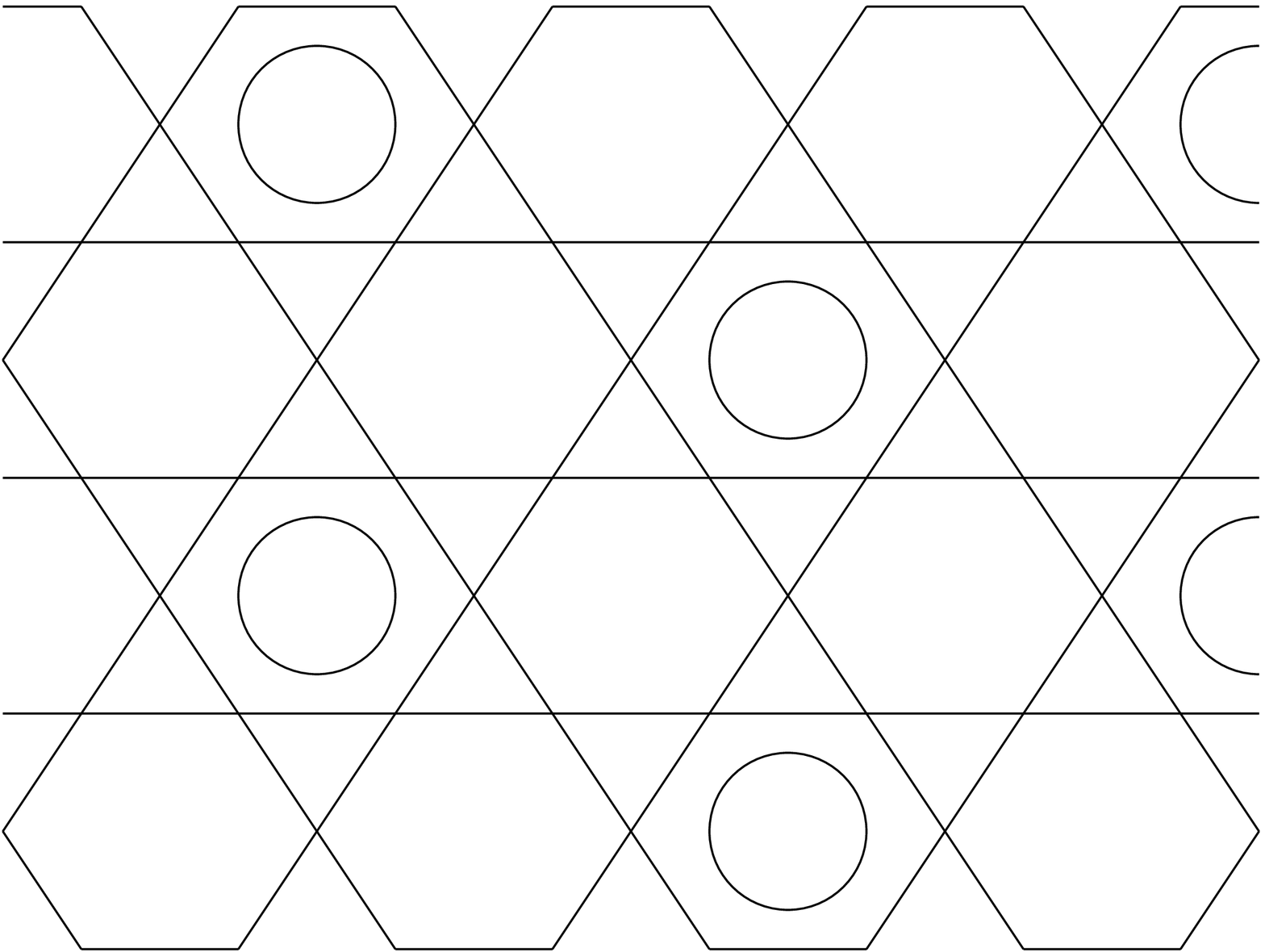,scale=0.21}} 
\end{center}
\caption{\label{fig3} Localized magnons on the \kagome lattice. Left panel:
One-magnon state indicated by a circle, the $+$ and $-$ signs 
correspond to the sign of the coefficients
$a_i=\pm 1$, see Eq. (\ref{eq6}). Right panel: Magnon crystal state
corresponding to the maximum filling $n_{max}=N/9$ of the \kagome
lattice with localized magnons (circles). 
The figures were taken from Ref. \cite{prl02}.  
}
\end{figure}

As an example we consider the HAFM (\ref{eq1}) on the 
\kagome lattice \cite{prl02}, i.e. we have $J_{ij}=1$ for nearest-neighbor
(NN) bonds and $J_{ij}=0$ else. 
Its one-magnon dispersion shown in 
Fig.~\ref{fig1} exhibits
one flat mode. The resulting localized magnon lives on a  
hexagon, see Fig.~\ref{fig3}, left panel. Its wave function is 
\be \label{eq6} 
|1\rangle^{\mbox{\tiny \kagome}}_L
 =\frac{1}{\sqrt{12s}} \sum_{i=1 \atop (i \in hexagon)}^6 a_i{\hat S}_i^-|0\rangle
\quad ,\quad a_i=(-1)^i \;.
\ee
Note that the number of hexagons $N/3$ corresponds to the 
number of states
in  the flat branch of $w({\bf k})$.
%
%
Because of the  localized nature of the  magnon we can put further such
magnons on the lattice such that there is no interaction between them.
The maximum filling $n_{max}$
of the \kagome lattice with localized magnons is shown in Fig.~\ref{fig3},
right panel. The resulting eigenstate is a  
magnon '{crystal}' state with $n_{max}=N/9$ magnons and 
a magnetic unit cell three times as large as
the geometric one.
Therefore the states with $n=\{0, 1, 2, 3, \ldots, N/9\}$  localized magnons
represent class of exactly known product eigenstates 
with quantum numbers 
$M=S_z= Ns - n=\{Ns, \;Ns-1, \;Ns-2,\;\ldots, \; Ns-N/9\}$. 

Due to the mutual
independence
of the localized magnons for $n \le n_{max} \propto N$
the energy $E_{loc}$ of the localized-magnon states 
is proportional to the number $n$ of localized magnons,
i.e. at $h=0$ one has $E_{loc}=E_{FM} - n \varepsilon_1$, 
where $E_{FM}$ is the energy of the fully polarized (vacuum) 
state and $\varepsilon_1$ is the energy  gain by one magnon. 
As a results  there is a simple linear relation between $E_{loc} $ and the
total magnetization $M$ valid for all systems hosting localized magnons  
\be
\label{eq7}
E_{loc}(M,h=0)
=-aN + bsM,
\ee 
where the parameters $a$ and $b$ depend on details of the system like
the exchange constant $J_{ij}$, the anisotropy parameter $\Delta$, 
the coordination number $z$ 
of the lattice etc.
For example for the isotropic spin-half HAFM with
NN exchange $J$
on the \kagome lattice one finds \cite{prl02,der04} 
\be \label{eq8}
E_{loc}(M,h=0)
=2s^2NJ-6snJ
=-4s^2NJ + 6sJM.
\ee
Note that also the spin-spin correlation functions of such
states can be easily found \cite{ri04}.

For the physical relevance of these eigenstates it is crucial  that they 
have  lowest energy in the  corresponding sector of $M$. Indeed this can
be proved for quite general anitferromagnetic spin models
\cite{schn01,hj_schmidt02}.

\begin{figure}[t!]
\centerline{\epsfig{file=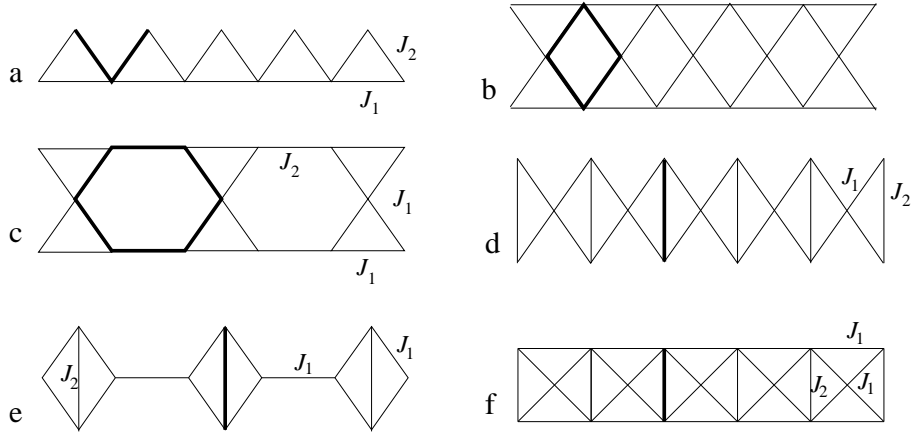,width=12cm}}
\caption{One-dimensional systems with localized-magnon states. The magnons
live on the restricted area  indicated by a thick line. 
a: sawtooth chain \cite{sawt,sawt1}, b: \kagome chain I  \cite{kago_cI}, c:
\kagome chain II \cite{kago_cII}, d: diamond chain  \cite{diamond}, 
e: dimer-plaquette chain  \cite{ivanov97}, f: 
frustrated ladder  \cite{ladder}. Note that there are special restrictions
for the exchange integrals to have localized-magnon states as lowest
eigenstates in the corresponding sector of $M$.}
\label{fig4}
\end{figure}

\begin{figure}[t!]
\hspace{-0.2cm}\epsfig{file=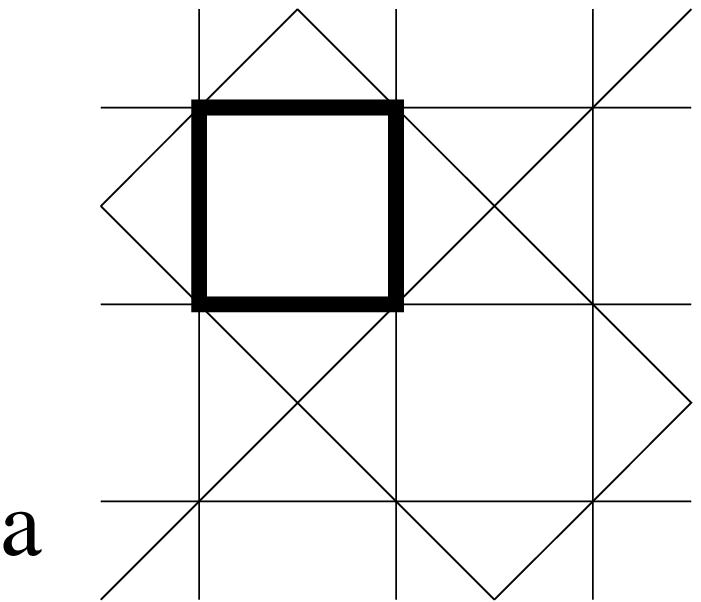,scale=0.5,angle=-0}
\hspace{+1cm}\epsfig{file=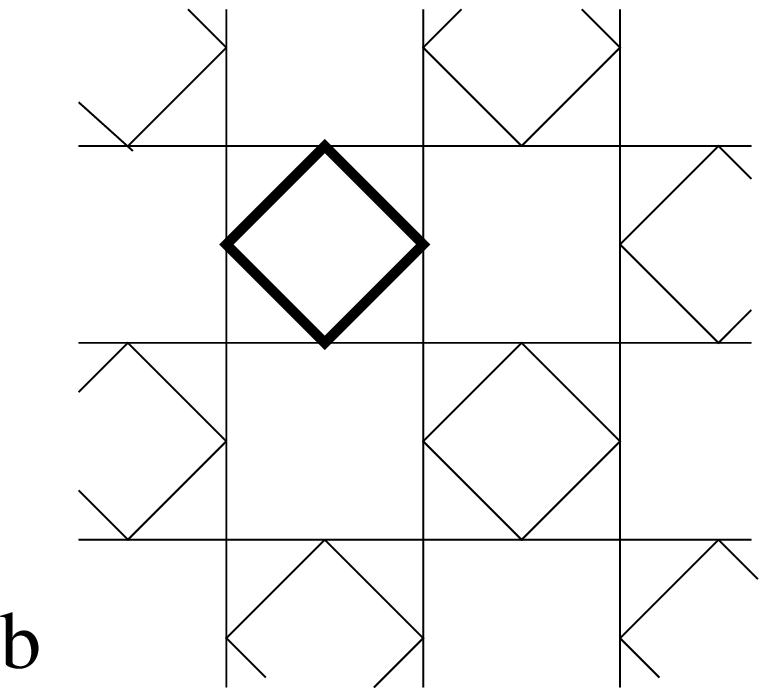,scale=0.48,angle=-0}
\hspace{+1cm}\epsfig{file=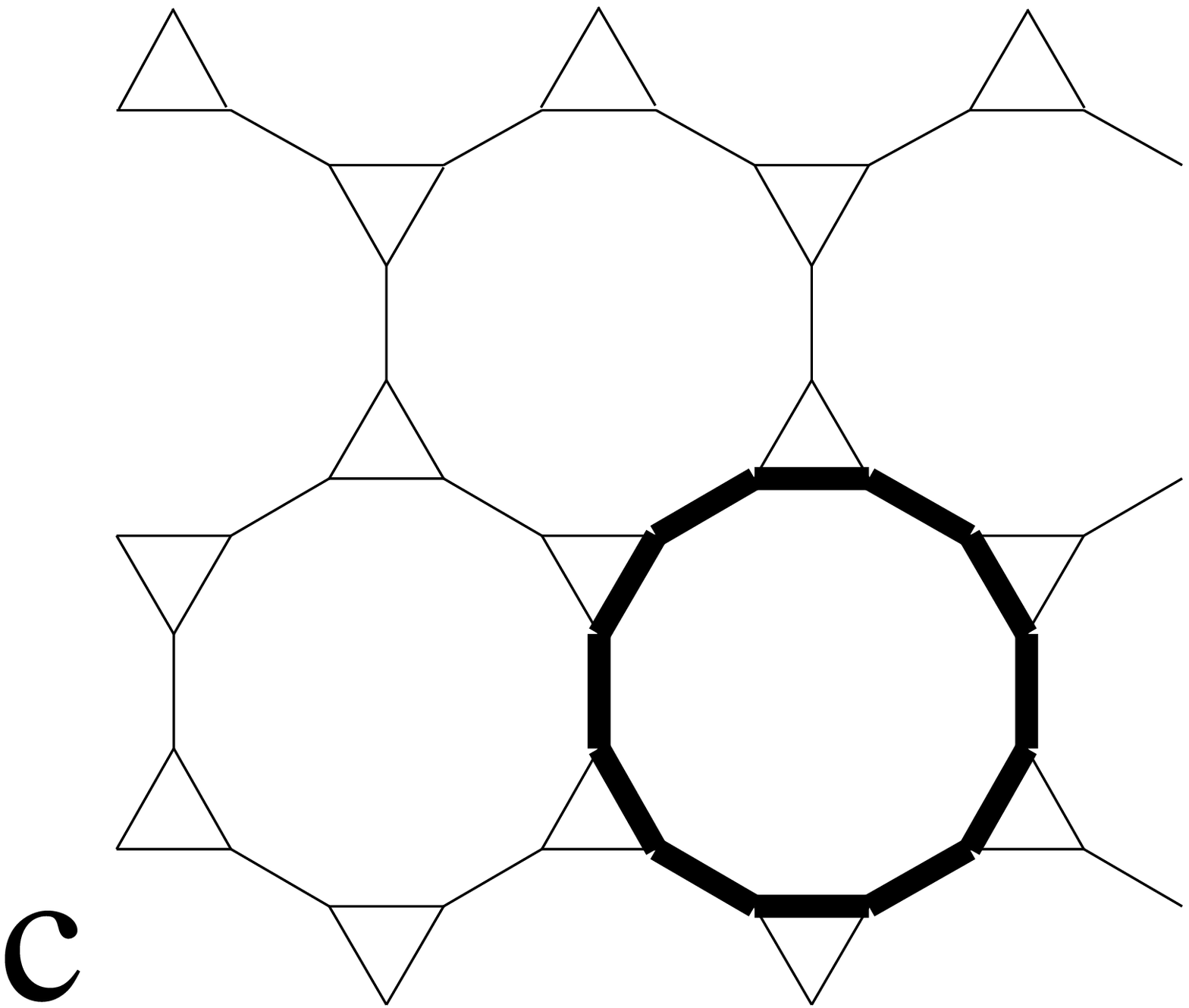,scale=0.24,angle=-0}

\hspace{-1cm}\epsfig{file=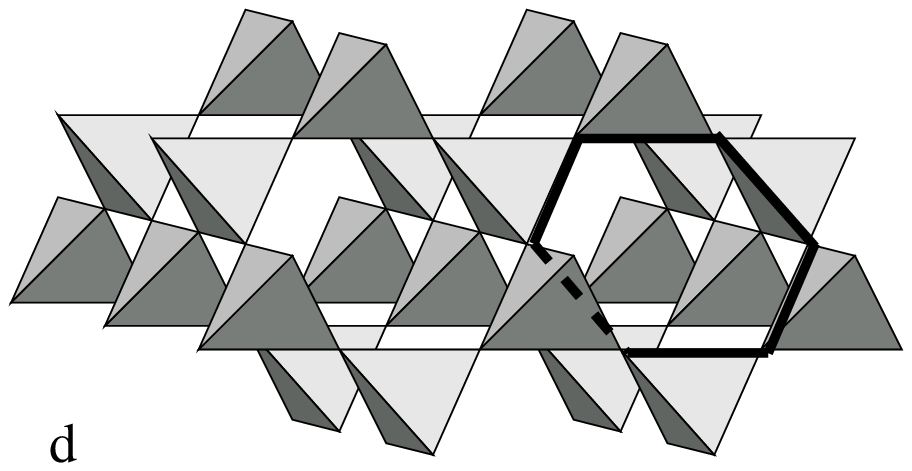,scale=0.8,angle=-0}
\caption{Two- and three-dimensional antiferromagnets 
with localized-magnon states. The magnons
live on the restricted area  indicated by a thick line. 
a: planar pyrochlore (checkerboard) lattice \cite{checker,ri04}, 
b: square-kagom\'e lattice \cite{sqago,sqago04}, c:
star lattice \cite{star04,Richter04}, d: pyrochlore  lattice \cite{canals98}. }
\label{fig5}
\end{figure}

Since the condition (\ref{eq4}) for the existence of localized-magnon states  
is quite general, one can find a lot of magnetic systems in one, two and
three dimensions having localized-magnon eigenstates 
\cite{schn01,prl02,ri04,sp04,der04,sawt1,star04,sqago04}. 
To illustrate that we show in Fig.~\ref{fig4} some 1D  and
in Fig.~\ref{fig5} some 2D and 3D antiferromagnetic spin lattices. 
But also for frustrated magnetic
molecules localized-magnon states are observed \cite{schn01}.
 
%

\section{Plateaus and jumps in the magnetization curve} \label{jump}
As discussed in the previous Section the localized-magnon states are the lowest
states in the corresponding sector of magnetization $M$. 
Hence they become ground states
in appropriate magnetic fields $h$. Furthermore we stated that  
there is a linear relation 
between the energy of these states $E_{loc}$ and $M$, cf. Eq. (\ref{eq7}). 
Applying a magnetic
field $h$ the energy reads $E_{loc}(M,h)=-aN
+ bsM - hM$ and  
one has a complete degeneracy
of all localized-magnon states
at the saturation field $h=h_{sat}=bs$.
As a result of this degeneracy
the zero-temperature magnetization $m=M/(Ns)$ jumps
between the saturation value $m=1$
and the value $1 - n_{max}/(Ns)$ corresponding to the maximum number
 $n_{max}$
of independent localized magnons.
Since  $n_{max}$ is proportional to $N$ but 
independent of the spin quantum number $s$, the 
height of the jump $\delta m=n_{max}/(Ns)$ goes to zero for $s \to \infty$, i.e.
the  magnetization jump due  to the localized-magnon states
becomes irrelevant if the spins become classical.

\begin{figure}
\begin{center}\epsfxsize=17.7pc  \epsfbox{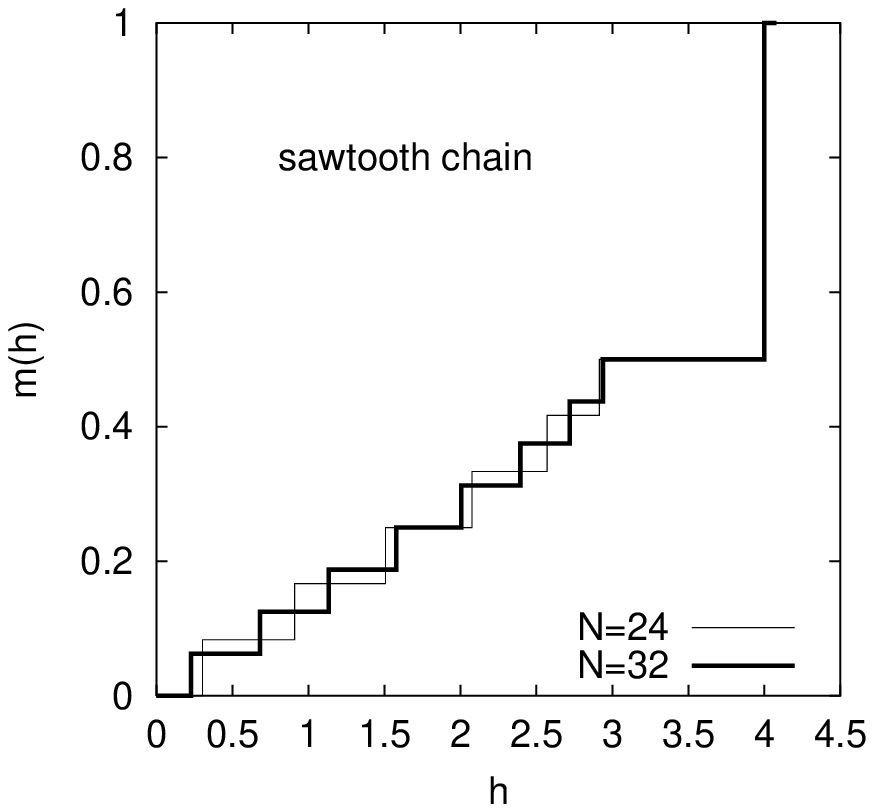}
\epsfxsize=19.7pc \epsfbox{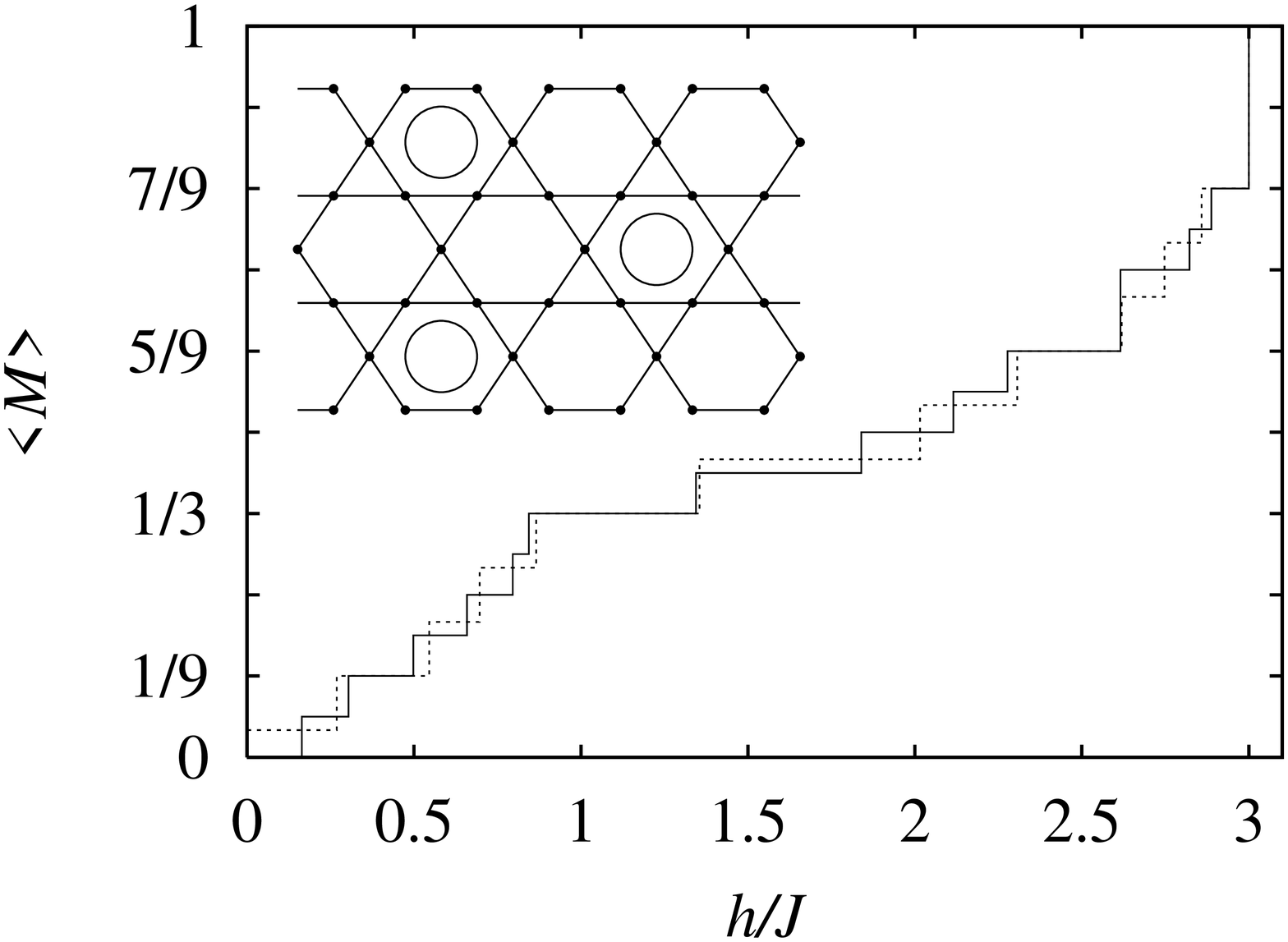}
\end{center}
\caption{\label{fig6}Magnetization  versus magnetic field $h$ of the
isotropic spin-half Heisenberg antiferromagnet.
Left panel:
The sawtooth chain with $J_1=1$ and $J_2=2J_1$, cf. Fig.~\ref{fig4}. 
The figure was  taken from Ref. \cite{ri04}.
Right panel:
The \kagome lattice  with $N=27$ (dashed line) and $36$ sites (solid line).
The inset illustrates the 'magnon crystal' state corresponding to
maximum filling with localized magnons, where the location of the 
 magnons is indicated by the
circles in certain hexagons.
The figure was taken from Ref. \cite{hon04}.
}
\end{figure}

We present in Fig.~\ref{fig6} two examples for the magnetization curves,
further examples can be found in
\cite{schul02a,Richter04,ri04,star04,sqago04}.
The jumps of height $\delta m= 1/2$ (sawtooth chain) 
and $\delta m= 2/7$ (kagom\'e lattice) are 
well pronounced.
Furthermore we see a wide plateau at the foot of the jump for the sawtooth
chain. Note that  
there are general arguments in favor of a plateau just below the jump 
\cite{momoi00,oshikawa00,Richter04}. Therefore we might expect a plateau
preceeding the jump in all magnetic systems with localized magnon states.
Though from Fig.~\ref{fig6} it remains
unclear whether there is a plateau for the \kagome lattice, too, a more
detailed analysis \cite{sp04} yields indeed evidence for a finite
plateau width of about $\delta h \sim 0.07J$.

\section{Magnetic-field induced spin-Peierls instability
       in quantum spin lattices with localized-magnon states}
The influence of a magnetoelastic coupling in frustrated antiferromagnets
on their low-temperature properties is currently widely discussed.
Lattice instabilities breaking the translational symmetry
are reported for 1D, 2D as well 3D quantum spin systems, see e.g. 
\cite{SP}.
In all those studies
the lattice instability was considered  at zero field.
As discussed already in an early paper by Gross  \cite{gross}
a magnetic field usually acts against the spin-Peierls transition
and might favor a uniform or incommensurate phase.
In contrast to those findings,
in this Section  we discuss a lattice instability in frustrated 
spin lattices hosting localized magnons
for which the magnetic field  is essential \cite{sp04}.

First we point out 
that due to the localized nature of the magnons
we have an inhomogeneous distribution
of NN spin-spin correlations
$\langle { \bf \hat S}_i {\bf \hat S}_j\rangle$ \cite{ri04}.
In case that 
one magnon is distributed uniformly over the lattice the deviation of the 
NN correlation from the ferromagnetic value, i.e. the quantity 
$\langle{\bf{ \hat S}}_i{\bf{\hat S}}_j\rangle-\frac{1}{4}$,
is of the order
$\frac{1}{N}$.
On the other hand
for a localized magnon (\ref{eq3})
we have along the polygon/line hosting the localized magnon
actually  in general a negative NN correlation
$\langle{\bf{\hat S}}_i{\bf{\hat S}}_j\rangle$  and all
other NN correlations are positive. For instance for the spin-half
HAFM on the \kagome lattice the localized-magnon state
(\ref{eq6}) leads to  
$\langle{\bf{\hat S}}_i{\bf{\hat S}}_j\rangle=-\frac{1}{12}$ for neighboring
sites $i$ and $j$ on a hexagon hosting a magnon 
and to  
$\langle{\bf{\hat S}}_i{\bf{\hat S}}_j\rangle=+\frac{1}{6}$ 
for neighboring
sites $i$ and $j$ on an attaching
triangle.
Hence
a deformation with optimal gain in magnetic energy
shall lead
to an increase of antiferromagnetic bonds
on the polygon/line  hosting the localized magnon 
(i.e. hexagon for the \kagome lattice) 
and to a decrease of the bonds on the attaching
triangles.

\begin{figure}
\begin{center}
\epsfxsize=14.pc
\hspace{-0.0cm}\epsfbox{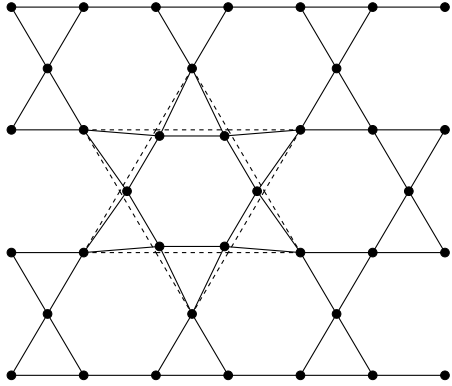}
\end{center}
\caption{
Kagom\'{e} lattice with one distorted hexagon
which can host localized magnons.
The parts of the lattices before distortions
are shown by dashed lines.
All bonds in the lattices before distortions have the same length.
The figure was taken from Ref. \cite{sp04}.
\label{fig7}}
\end{figure}

We will discuss the situation for the isotropic spin-half HAFM
on the
\kagome lattice in more detail.
A corresponding deformation 
 which preserves the symmetry of the cell
which hosts the localized magnon
is 
shown in Fig.~\ref{fig7}.
To check the  stability of the \kagome  lattice
with respect to a spin-Peierls mechanism we must compare the magnetic and
the elastic energies. 
For the kagom\'{e} lattice  the
deformation shown in Fig.~\ref{fig7}  
leads to the following changes in the exchange interactions:
$J\to \left(1+\delta\right)J$
(along the edges of the hexagon)
and
$J\to \left(1-\frac{1}{2}\delta\right)J$
(along the two edges of the triangles attached to the hexagon),
where the quantity
$\delta$
is proportional to the displacement of the atoms and the change in the
exchange integrals due to lattice distortions is taken into account
in first order in $\delta$.
The magnetic energy (\ref{eq8})
is lowered by distortions
and becomes for one magnon and one corresponding distortion 
$E_{loc}(n=1,h=0,\delta)=\frac{1}{2}NJ-3J
-\frac{3}{2}\delta J$. Considering $n \le n_{max}$ independent
localized magnons and  corresponding distortions the energy gain is then 
$e_{mag}= - n \frac{3}{2}\delta J$.
 On the other hand the elastic energy
in harmonic approximation increases according to $e_{elast} \propto
\delta^2$. 
Therefore 
a minimal total energy is obtained for
a finite $\delta=\delta^\star > 0$.
For the \kagome lattice the elastic energy for one distorted cell is 
 $e_{elast}= 9\gamma\delta^2$
(the parameter
$\gamma$
is proportional to the elastic constant of the lattice).
If the
localized-magnon states are the ground states of the systems then we have a
favorable 
spin-Peierls distortion  with $\delta^\star=\frac{1}{12}
\frac{J}{\gamma}$.
As discussed in the previous Section we expect a plateau at the foot of the
magnetization jump. The spin-Peierls distortion due to localized-magnon states 
  then takes place for the values of the magnetic field belonging to this
plateau.

Now the question arises
whether the lattice distortion under consideration
is stable  below and above this plateau,
i.e., for
$M < \frac{1}{2}N-n_{max}$ and $M = \frac{1}{2}N$.
It is easy to check that the lattice distortion illustrated in
Fig.~\ref{fig7}  
is not favorable for the fully
polarized vacuum state, i.e. for $M=\frac{1}{2}N$.
For magnetizations $M$ below this plateau
we are not able to give a rigorous answer
but  numerical results 
for finite \kagome lattices
of size $N=18, 27, 36, 45, 54$ indicate that there is no spin-Peierls
deformation
adopting the lattice distortion shown in Fig.~\ref{fig7}
for $M < \frac{1}{2}N-n_{max}$ (for more details, see \cite{sp04}).

We mention that the scenario discussed above  basically remains unchanged
for the anisotropic Hamiltonian (\ref{eq1})
with $\Delta\ne 1$ and also for spin quantum number $s > 1/2$ \cite{sp04}.

From the experimental point of view
the discussed effect
should most spectacularly manifest itself
as a hysteresis in the magnetization and the deformation
of kagom\'{e}-lattice antiferromagnets
in the vicinity of the saturation field.
We emphasize that the discussed 
spin-Peierls instability in high magnetic fields
may appear in the whole class 
of frustrated quantum magnets in one, two and three dimensions
hosting independent localized magnons
provided it is possible to construct
a lattice distortion preserving the symmetry of the localized-magnon cell.

\section{Finite low-temperature entropy and enhanced magnetocaloric effect
       in the vicinity of the 	saturation field}\label{entro}

It is well-known that strongly  frustrated Ising or classical Heisenberg spin
systems may exhibit a huge ground-state degeneracy, see e.g.
\cite{lhuillier03,Richter04,wannier,moessner01,moes01,zhito,uda}.
For {\bf quantum} systems due to fluctuations
often the degeneracy found for classical systems
is lifted and the quantum ground state is unique (so-called '{\it order from
disorder'} phenomenon \cite{villain,shender}).
However,
highly frustrated antiferromagnetic quantum spin lattices
hosting localized magnons
are an example where one finds a huge ground-state degeneracy in a {\bf
quantum}
system.
As discussed in Section \ref{jump} the localized-magnon states
become degenerate at saturation field.
As pointed out first in \cite{Richter04}
the degeneracy 
grows exponentially with system size $N$.
In more detail this huge degeneracy and its consequence for the low
temperature physics were discussed in \cite{zhito04,zhito04a,der04}.

\begin{figure}
\begin{center}
\epsfxsize=22.0pc \epsfbox{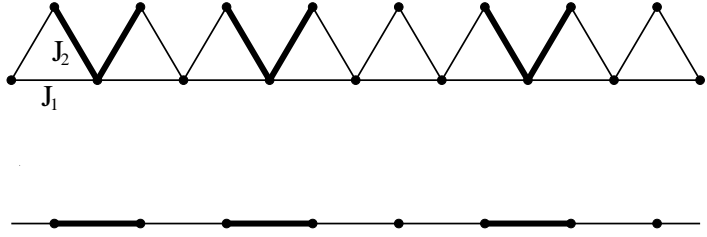}
\end{center}
\caption{\label{fig8}
The sawtooth chain which hosts three localized magnons
at fat {\sf{V}} parts
(top)
and the auxiliary lattice used for the exact calculation of 
the ground-state degeneracy at saturation
(bottom).
The localized magnons are eigenstates 
for $J_2 = \sqrt{2\left(1+\Delta\right)} J_1$ \cite{prl02}. 
The figure was taken from Ref. \cite{der04}.
}
\end{figure}

For some of the spin systems hosting localized magnons 
the ground-state degeneracy at saturation and therefore the residual 
entropy at zero temperature
can be calculated exactly by mapping the localized-magnon problem 
onto a related  lattice gas model of hard-core objects
\cite{zhito04,zhito04a,der04}.  
These lattice gas models of hard-core objects have been  studied 
over the last  decades in great detail 
(see, e.g. \cite{lat_gas}).
Let us illustrate  this for the sawtooth chain.
Here the local areas where the  magnons can live are the 'valleys'
of the sawtooth chain, see Fig.~\ref{fig4}a. 
Because a certain local area of the lattice can be occupied by a 
magnon or not, 
the degeneracy of the ground state at saturation,
${\cal{W}}$,
grows exponentially with $N$ 
giving rise to a finite zero-temperature entropy per site at saturation
\begin{eqnarray}
\label{02}
\frac{{\cal{S}}}{N}=\lim_{N\to\infty}\frac{1}{N}\log{\cal{W}} > 0.
\end{eqnarray}
Now we map
the original lattice which hosts  localized magnons 
onto an auxiliary lattice 
which is occupied by hard-core objects, which are 
rigid monomers and dimers in the case of the sawtooth chain, see 
Fig.~\ref{fig8}.
The auxiliary chain (Fig.~\ref{fig8}, bottom)
consists of ${\cal{N}}=\frac{1}{2}N$ sites 
which may be filled 
either by rigid monomers 
or by rigid dimers occupying two neighboring sites.
The limiting behavior of ${\cal{W}}$ for 
${\cal{N}}\to\infty$
was found by Fisher many years ago \cite{lat_gas}
\begin{eqnarray}
\label{04}
{\cal{W}}
=\exp\left(\log\frac{1+\sqrt{5}}{2}\;{\cal{N}}\right)
\approx
\exp\left(0.240606 N\right) \quad \to \quad 
\frac{{\cal{S}}}{N}=0.240606.
\end{eqnarray}
The relevance of this result
for experimental studies emerges at low but finite temperatures.
In Ref. \cite{zhito04} this mapping was used to obtain a quantitative
description of the low-temperature magnetothermodynamics in the vicinity of
the saturation field of the quantum antiferromagnet on
the \kagome lattice and on the sawtooth chain.  

\begin{figure}
\begin{center} \epsfig{file=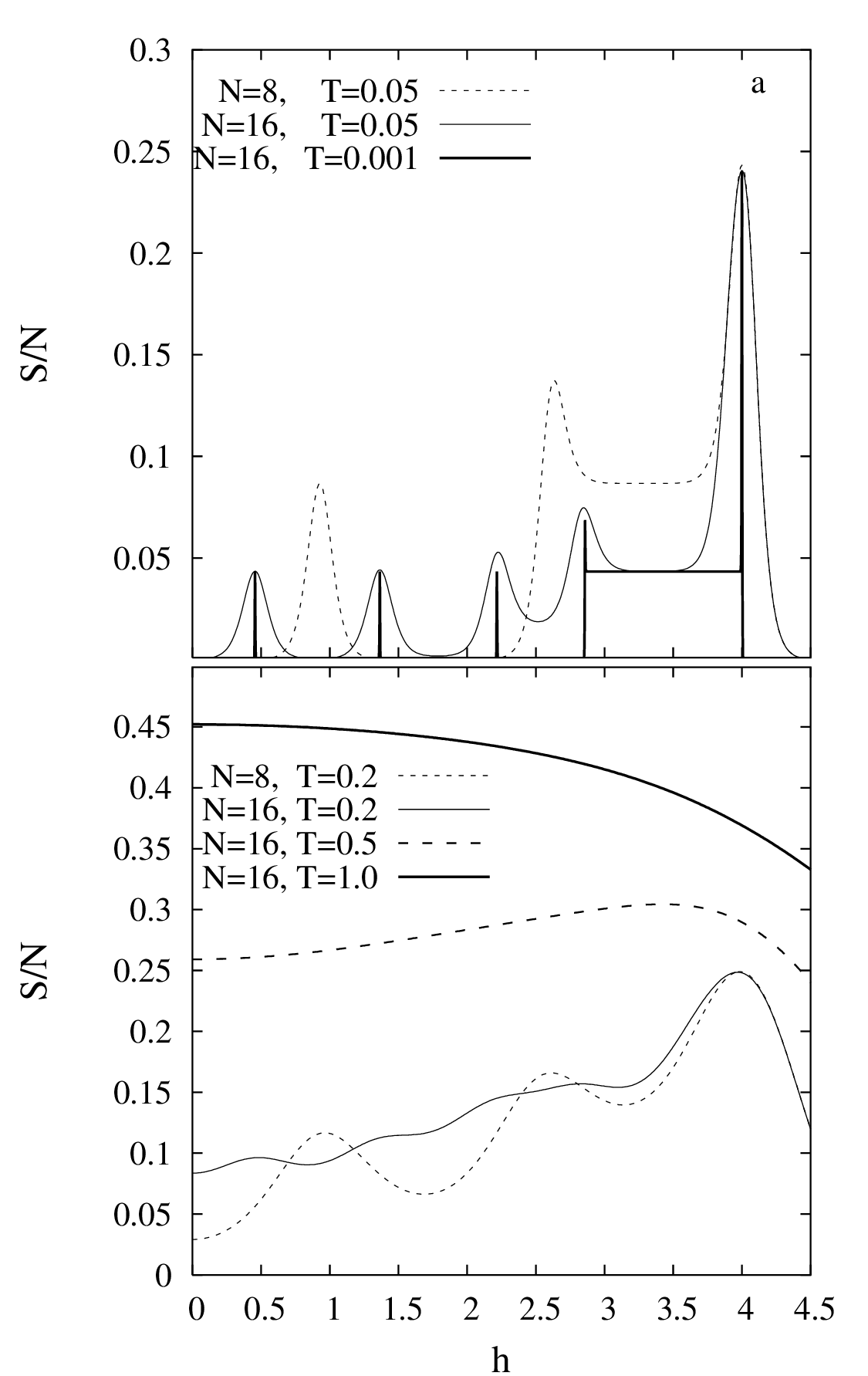,scale=0.4,angle=-90}
\epsfig{file=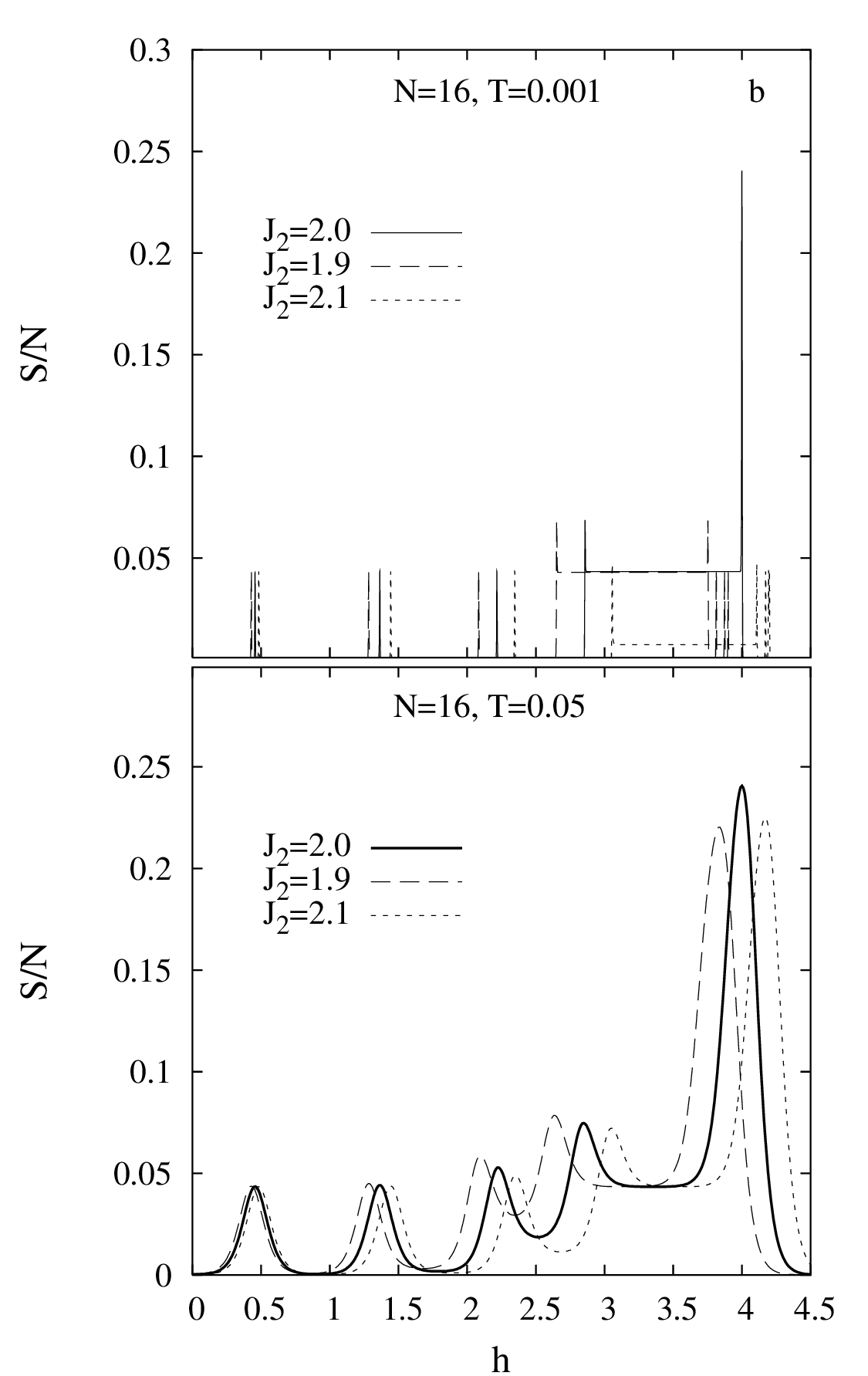,scale=0.4,angle=-90}
\end{center}
\caption{\label{fig9}
Field dependence of the isothermal entropy per site 
for the Heisenberg antiferromagnet on the  sawtooth chain 
of different length
($s=\frac{1}{2}$, 
$\Delta=1$).
a: $J_1=1$, $J_2=2$, i.e. the condition for bond strengths (\ref{eq4}) is
fulfilled and the localized-magnon states are exact eigenstates.
b: Influence of a deviation from the perfect condition for bond strengths.
The figures were taken from Ref. \cite{der04}.
}
\end{figure}

In what follows  we report  
on the extension of 
the analytical findings for the  zero-temperature entropy at the saturation
field to finite temperatures and 
to arbitrary magnetic fields
using exact diagonalization data 
for the sawtooth chain of up to $N=20$ sites \cite{der04,zhito04a}.
In Fig.~\ref{fig9}a we show the isothermal entropy versus magnetic field 
for several temperatures.
The presented results show
that for several magnetic fields below saturation $h<h_{sat}$ 
one has a two-fold or even a three-fold degeneracy of the energy levels 
leading in a finite system to a finite zero-temperature  entropy per site.
Correspondingly  one finds in Fig.~\ref{fig9}a (upper panel)
a peaked structure and moreover a plateau-like area 
just below $h_{sat}$. However, it
is clearly seen in Fig.~\ref{fig9}a (upper panel)
that the height of the peaks and of the plateau decreases with system size $N$
leading to $\lim_{N\to\infty}{{\cal{S}}/N}=0$ at $T=0$  for $h<h_{sat}$ and
$h>h_{sat}$. 
Only the peak at $h=h_{sat}=4$ is independent of $N$ and remains finite for $N
\to \infty$.
At finite but low temperatures this peak survives as a well-pronounced 
maximum and it only disappears if the temperature is of the 
order of the exchange constant, see Fig.~\ref{fig9}a (lower panel).
Note that the value of the entropy at saturation, 
which agrees with the analytical prediction (\ref{04}),
is almost temperature independent up to about $T \approx 0.2$.
Thus, 
the effect of the independent localized magnons 
leading to a finite residual  zero-temperature
entropy
is present  at finite temperatures $T \lesssim 0.2$
producing a noticeable maximum in the isothermal entropy curve 
at the saturation field.
We mention that the
numerical results for higher spin quantum numbers $s$ suggest 
that the enhancement of the entropy at saturation 
for finite temperatures becomes less pronounced with increasing $s$.

With respect to the  experimental observation 
of the maximum in the low-temperature entropy at the saturation field 
in real compounds we are  faced with the situation that 
the condition  on bond strengths, see Eq. (\ref{eq4}),  
under which the localized-magnon states become the exact eigenstates
are certainly
not strictly fulfilled.
For the considered 
isotropic spin-half HAFM on the  sawtooth chain 
this condition is fulfilled  for $J_2=2J_1$, see Fig.~\ref{fig9}a.
Based on the numerical calculations one is able
to discuss  the ``stability'' of the maximum in the entropy 
against deviation from the perfect condition for bond strengths.
In Fig.~\ref{fig9}b
we show the field dependence of the entropy at low temperatures
for the sawtooth chain of $N=16$ sites
with $J_1=1$ and $J_2=1.9$ and $J_2=2.1$. 
Since the degeneracy of the ground state at saturation is lifted
when $J_2 \ne 2J_1$,
the entropy at saturation at very low temperatures
(long-dashed and short-dashed curves 
in the upper panel of Fig.~\ref{fig9}b) is not enhanced at the saturation
field.
However,
the initially degenerate energy levels remain close to each other,
if $J_2$ only slightly deviates from the perfect value $2J_1$. 
Therefore
with increasing temperature  
those levels become accessible for the spin system
and one obtains again a maximum in the  entropy 
in the vicinity of saturation
at low but nonzero temperatures, see lower panel in Fig.~\ref{fig9}b
(long-dashed and short-dashed peaks in the vicinity of saturation).
We emphasize that the low-temperature maximum  of ${\cal{S}}/N$
at saturation 
is a generic effect for strongly frustrated quantum 
spin lattices 
which may host independent localized magnons.

Let us remark
that the ground-state degeneracy problem 
of antiferromagnetic Ising lattices in the critical magnetic field 
(i.e. at the spin-flop transition point),
which obviously do not contain quantum fluctuations,  
has been discussed in the literature, see e.g. \cite{metcalf}.

It has been  pointed out very recently by Zhitomirsky and Honecker
\cite{zhito03,zhito04a} that the most spectacular effect accompanying a 
maximum in the isothermal entropy ${\cal{S}}(h)$ is an enhanced magnetocaloric
effect.
Indeed the cooling rate for an adiabatic (de)magnetization process  
is proportional to the derivative of the isothermal entropy with respect to
the magnetic field 
\be
\left(\frac{\partial T}{\partial h}\right)_{\cal S} = -T \frac{(\partial
{\cal S}/\partial h)_T}{C},
\ee
where $C$ is the specific heat.
Again one can calculate the field dependence of the temperature for an 
adiabatic (de)magnetization process  
for finite systems by exact diagonalization.
Some results for the 
isotropic spin-half HAFM on the  sawtooth chain 
are shown in Fig.~\ref{fig10}. The lowest curves in Fig.~\ref{fig10} belongs
to ${\cal S}/N=0.05$ and $N=12,16,20$, respectively. The other curves correspond to 
${\cal S}/N=0.1, 0.15, \ldots, 0.4,  0.45$.
The magnetocaloric effect is largest in the vicinity of the saturation field.
In particular, a demagnetization coming from magnetic fields larger than
$h_{sat}$  is very efficient. If one starts for $h > h_{sat}$ with an entropy
lower than the residual entropy at $h_{sat}$,  ${\cal S}/N=0.240606$ 
(see Eq. ({\ref{04})),
then one observes even a cooling to $T \to 0$ as $h \to h_{sat}$. Thus
frustrated magnetic systems hosting localized magnons  allow
magnetic cooling from quite large $T$ down to very low temperatures. 
We mention, that the results shown in Fig.~\ref{fig10} also 
clearly demonstrate, that
finite-size effects are very small for $h \gtrsim h_{sat}$ at any temperature.
Therefore the above discussion is valid also for large systems $N \to \infty$.
\begin{figure}
\begin{center} \epsfig{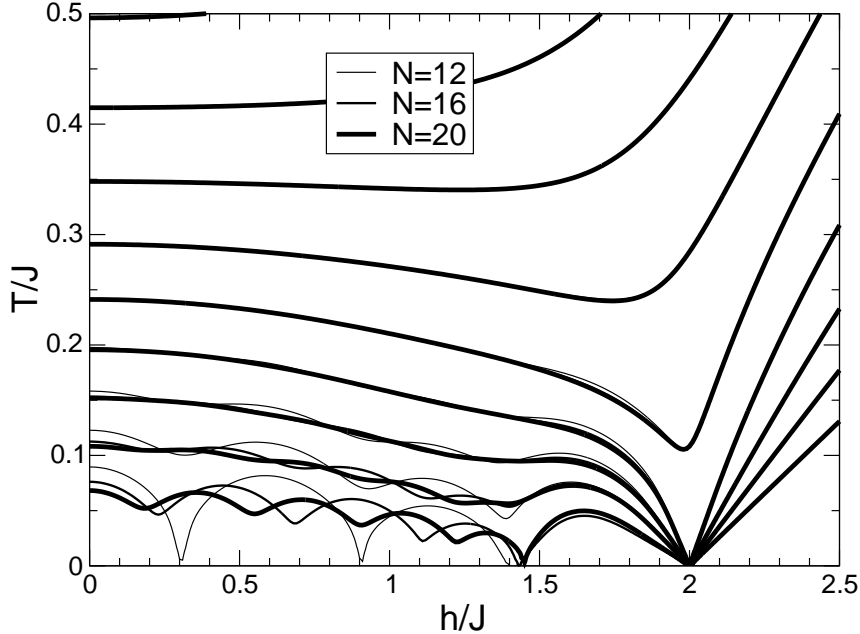}
\end{center}
\caption{\label{fig10}
Lines of constant entropy versus field (i.e. adiabatic (de)magnetization
curves) 
for the Heisenberg antiferromagnet on the sawtooth chain of different length
($s=\frac{1}{2}$, 
$\Delta=1$,
$J_1=J/2$, $J_2=J$).
The figure was taken from Ref. \cite{zhito04a} with friendly permission of
A.~Honecker.
}
\end{figure}

\section{Summary}
We have reviewed recent results 
 \cite{schn01,prl02,ri04,sp04,zhito04,zhito04a,der04,star04,sqago04}
 on exact eigenstates constructed from
localized magnons which appear in a class of  
 frustrated spin lattices. For these eigenstates several quantities like the
energy and the spin-spin correlation can be calculated analytically. 
The physical relevance of the  localized-magnon  eigenstates emerges at high
magnetic fields where they can become ground states of the spin system.

For frustrated magnetic systems having localized-magnon ground states several
interesting physical effects associated with this states may occur.
First one   finds 
a macroscopic jump in the
zero-temperature magnetization curve at the saturation field $h_{sat}$.
This jump is a true quantum effect which vanishes if the spins become classical
($s\rightarrow\infty$).
At the foot of the jump one can expect a plateau in the magnetization curve.

Since all localized-magnon states have the same energy  at $h=h_{sat}$
 a  huge degeneracy of the ground state at saturation 
is observed which increases exponentially
with system size $N$ thus    
leading to a nonzero residual  zero-temperature entropy.
For some of the frustrated spin models hosting localized magnons 
the residual entropy at saturation field
can be estimated exactly.
At finite temperatures $T$   
the localized-magnon states 
produce a maximum in the isothermal  entropy versus field curve  
in the vicinity of the saturation field for not too large $T$. 
This maximum in the isothermal entropy at $h_{sat}$ 
leads to 
an enhanced magnetocaloric
effect.
If one starts for $h > h_{sat}$ with an entropy
lower than the residual entropy at $h_{sat}$ 
then one observes even a cooling to $T \to 0$ as $h \to h_{sat}$. This may 
allow
cooling from quite large $T$ down to very low temperatures.

Last  but not least 
the localized-magnon states may  lead to  
a spin-Peierls  instability in strong magnetic fields,
for instance 
for the antiferromagnetic \kagome spin lattice.
For this system   
the magnetic-field driven spin-Peierls instability
breaks spontaneously the translational symmetry of the kagom\'{e}
lattice
and appears only in 
a certain region of the magnetic field near saturation.

We emphasize that the reported effects are
generic in highly frustrated magnets. To observe them in experiments one needs
frustrated magnets with small spin quantum number $s$ and sufficiently
small exchange coupling strength $J$ to reach  the saturation field.
There is  an increasing number of synthesized quantum frustrated spin 
lattices,
e.g. quantum antiferromagnets with  
a kagom\'{e}-like
structure \cite{ramirez00dd,19,20,bono04}.
Though such   materials often do not fit  
perfectly to the  lattice geometry having 
localized-magnon goundstates, 
the physical effects based on localized
magnons states may survive in non-ideal geometries (see 
Section \ref{entro}), which may open the window to the experimental 
observation 
of the theoretically predicted effects.

\vspace{0.5cm}

{\bf Acknowledgment}\\
The author is indebted to O.~Derzhko, A.~Honecker, 
H.-J.~Schmidt, J.~Schnack, and J.~Schulenburg
for the fruitful collaboration in the field of localized-magnon 
states in frustrated antiferromagnets. Most of the results discussed
in the paper are based on common publications with these colleagues.
In particular, I thank J.~Schulenburg for many valuable discussions and
hints.
Furthermore I thank H.~Frahm for directing my attention
to the flat-band ferromagnetism. 
I also acknowledge the support from the Deutsche Forschungsgemeinschaft
(project No. Ri615/12-1).


\end{document}